\documentclass[lettersize,journal]{IEEEtran}
    \usepackage{booktabs}
    \usepackage{diagbox}
    \usepackage{makecell}
    \usepackage{array}
    \usepackage{graphicx,amssymb,amsmath}
    \usepackage{multicol}
    \usepackage[noadjust]{cite} 
    \usepackage{setspace}
    \usepackage{subfigure}
    \usepackage{graphicx}
    \usepackage{float}
    \usepackage {url}
    \usepackage{stfloats}
    \usepackage{amsthm,pifont}
    \usepackage{flushend}
    \usepackage{cases,subeqnarray}
    \usepackage{bm,multirow,bigstrut}
    \usepackage{amsmath, amsthm, amssymb}
    \usepackage{textcomp}
    \usepackage{latexsym,bm}
    \usepackage{booktabs}
    \usepackage{mathtools}
    \usepackage{dsfont}
    \usepackage{extarrows}
    \usepackage{epsfig}
    \usepackage{epsfig}
    \usepackage{epstopdf}
    \usepackage[table]{xcolor}
    \usepackage{colortbl} 
    \usepackage{multirow, hhline}
    \usepackage{xcolor}
    \usepackage{array} 
    \usepackage{algorithmicx,algorithm}
    \usepackage{hyperref}

    \theoremstyle{plain}

    \theoremstyle{plain}

    \usepackage{amsmath}

    \IEEEoverridecommandlockouts
    \begin{document}
    \title{Generative AI for Lyapunov Optimization Theory in UAV-based Low-Altitude Economy Networking}
    
    \author{
    Zhang Liu,
    Dusit Niyato,~\IEEEmembership{Fellow,~IEEE},
    Jiacheng Wang,
    Geng Sun,
    Lianfen Huang, \\Zhibin Gao, and Xianbin Wang,~\IEEEmembership{Fellow,~IEEE}
    \thanks{The work was supported in part by the National Natural Science Foundation of China under Grant 62371406 and Grant 62171392, in part by the Natural Science Foundation of Xiamen, China under Grant 3502Z202473053, in part by the Key Science and technology Project of Fujian Province under Grant 2024H6030 and 2021J01004, in part by the National Research Foundation, Singapore, and in part by Infocomm Media Development Authority under its Future Communications Research \& Development Programme, Defence Science Organisation (DSO) National Laboratories under the AI Singapore Programme under Grant FCP-NTU-RG-2022-010 and Grant FCP-ASTAR-TG-2022-003, in part by Singapore Ministry of Education (MOE) Tier 1 under Grant RG87/22, and in part by the NTU Centre for Computational Technologies in Finance (NTU-CCTF). \emph{ (Corresponding author: Lianfen Huang).}}
    \thanks{Z.~Liu and L.~Huang are with the Department of Informatics and Communication Engineering, Xiamen University, Fujian 361102, China. (e-mail: zhangliu@stu.xmu.edu.cn, lfhuang@xmu.edu.cn).}
    \thanks{D.~Niyato and J.~Wang are with the College of Computing and Data Science, Nanyang Technological University, Singapore. (e-mail: dniyato@ntu.edu.sg, jiacheng.wang@ntu.edu.sg).}
    \thanks{G.~Sun is with the College of Computer Science and Technology, Jilin University, Changchun 130012, China. (e-mail: sungeng@jlu.edu.cn).}
    \thanks{Z. Gao is with the Navigation Institute, Jimei University, Xiamen, Fujian 361021, China. (gaozhibin@jmu.edu.cn).}
    \thanks{X.~Wang is with the Department of Electrical and Computer Engineering, Western University, London, Ontario N6A 5B9, Canada. (xianbin.wang@uwo.ca).}
    }
\maketitle
    \begin{abstract}
        Lyapunov optimization theory has recently emerged as a powerful mathematical framework for solving complex stochastic optimization problems by transforming long-term objectives into a sequence of real-time short-term decisions while ensuring system stability. This theory is particularly valuable in unmanned aerial vehicle (UAV)-based low-altitude economy (LAE) networking scenarios, where it could effectively address inherent challenges of dynamic network conditions, multiple optimization objectives, and stability requirements. Recently, generative artificial intelligence (GenAI) has garnered significant attention for its unprecedented capability to generate diverse digital content. Extending beyond content generation, in this paper, we propose a framework integrating generative diffusion models with reinforcement learning to address Lyapunov optimization problems in UAV-based LAE networking. We begin by introducing the fundamentals of Lyapunov optimization theory and analyzing the limitations of both conventional methods and traditional AI-enabled approaches. We then examine various GenAI models and comprehensively analyze their potential contributions to Lyapunov optimization. Subsequently, we develop a Lyapunov-guided generative diffusion model-based reinforcement learning framework and validate its effectiveness through a UAV-based LAE networking case study. Finally, we outline several directions for future research.
    \end{abstract}
    \begin{IEEEkeywords}
    Generative AI, Lyapunov optimization theory, unmanned aerial vehicle, low-altitude economy networking.
    \end{IEEEkeywords}
    \IEEEpeerreviewmaketitle
    \section{Introduction}
    Lyapunov optimization theory is a mathematical framework originating from Lyapunov drift theory~\cite{cui2012survey}, commonly studied in fields such as communication networks and control systems. It has been widely used in dynamic environments to stabilize and optimize queueing systems under stochastic conditions. For example, in wireless networks, bandwidth resources can be dynamically allocated based on the length of the data queue measured by a Lyapunov function, ensuring continuous data flow and maximizing throughput. In smart grids, electrical loads can be scheduled by minimizing the Lyapunov drift, thereby reducing energy consumption while guaranteeing power system stability~\cite{7845698}. The widespread application of Lyapunov optimization theory underscores its value in optimizing system performance in dynamic and uncertain environments.

    In dynamic unmanned aerial vehicle (UAV) networks with limited power supply, UAVs must continuously adjust their positions and transmission power to match fluctuating channel conditions while managing long-term energy consumption constraints. These optimization decisions are further complicated by the need for computation offloading within the constraints of available computational resources. Additionally, the complexity intensifies in low-altitude economy (LAE) networking\footnote{\href{https://www.reuters.com/world/china/china-test-flies-biggest-cargo-drone-low-altitude-economy-takes-off-2024-08-12/}{https://www.reuters.com/world/low-altitude-economy}}, which supports a large number of UAVs while ensuring strict quality of service (QoS), security, and safety requirements. Therefore, Lyapunov optimization theory is desirable for LAE networks, as it enables the dual benefits of optimizing system performance and ensuring operational stability without requiring knowledge of future environmental information~\cite{9687317}. A substantial body of literature has explored methods for addressing the Lyapunov optimization problem, primarily focusing on traditional approaches such as convex optimization and heuristic algorithms. However, these conventional methods often face distinct challenges.

    \begin{itemize}
\item {\textbf{High Computational Complexity.}} Convex optimization methods rely on extensive computations, making it challenging to meet the real-time decision-making requirements in dynamic LAE networks. Additionally, Lyapunov optimization problems in high-dimensional LAE networks often involve non-convexity, rendering convex optimization methods ineffective.


\item {\textbf{Limited Generalizability.}} Heuristic algorithms are typically designed based on specific problem features and are difficult to generalize across different environments, limiting their applicability in dynamic LAE networks. Furthermore, heuristic algorithms yield only sub-optimal solutions, without guaranteeing global optimality~\cite{7274642}. 

\item {\textbf{Dependency on Complete Information.}} Conventional methods require precise knowledge of the system state to formulate and solve the optimization problem. However, such complete information is often unavailable in practical LAE networks, where channel conditions, user locations, and network demands change rapidly~\cite{wu2021survey}.

\end{itemize}

Several works have explored learning-based methods, such as supervised learning and reinforcement learning, to address the Lyapunov optimization problem. For instance,  supervised learning uses labeled data (system state-optimal decision pairs) to train neural networks, while reinforcement learning incorporates the Lyapunov drift-plus-penalty function into its reward function, enabling agents to learn optimal policies through continuous interaction with the environment. However, these learning-based approaches also feature specific limitations below.

\begin{itemize}
\item {\textbf{Reliance on High-Quality Annotated Data.}} Supervised learning requires large amounts of high-quality annotated data, often necessitating expert strategies. In real-time LAE networks, covering a wide range of annotated data can be challenging, particularly in dynamic environments where annotations can quickly become invalid~\cite{ayers2023supervised}.

\item {\textbf{Difficulty in Balancing Exploration and Exploitation.}} Reinforcement learning improves policies through exploration; however, excessive exploration can lead to sub-optimal solutions. On the other hand, excessive exploitation can result in short-sightedness, neglecting the long-term optimality of LAE networks~\cite{liu2024dnn}.
\end{itemize}

Fortunately, generative artificial intelligence (GenAI) techniques, such as variational autoencoders (VAEs), generative adversarial networks (GANs), and generative diffusion models (GDMs) offer promising solutions to address the aforementioned challenges. For example, VAEs can compress high-dimensional states into low-dimensional latent variables, significantly reducing computational complexity. GANs can produce large amounts of high-quality synthetic data with specific labels. GDMs can improve the balance between exploration and exploitation by adjusting the number of denoising steps. Compared to conventional methods and traditional AI, the advantages of using GenAI to solve the Lyapunov optimization problem in LAE networks include the following:

\begin{itemize}
\item {\textbf{Training Data Generation.}} Sudden extreme states, such as network congestion or abrupt channel deterioration, are critical points of instability in LAE networks. GenAI can generate corresponding synthetic data to minimize the Lyapunov drift-plus-penalty function, thereby reducing the risk of instability during actual operations.

\item {\textbf{Latent Space Representation.}} GenAI can map high-dimensional LAE network states into a low-dimensional latent space. This transformation allows resource allocation and control strategies that minimize the Lyapunov drift-plus-penalty function to be identified more efficiently, thereby reducing computational burden.

\item {\textbf{Computational Efficiency.}} GenAI's ability to rapidly generate candidate strategies makes it computationally efficient for balancing Lyapunov drift and penalty terms in dynamic environments. This capability is critical for LAE networks, where real-time changes in network state demand a fast response to ensure stability.
\end{itemize}

Motivated by these, this paper aims to provide a forward-looking exploration into the integration of GenAI and Lyapunov optimization theory for LAE network optimization. The contributions of this paper are summarized as follows:

\begin{itemize}
\item We begin with an overview of Lyapunov optimization theory, examining the limitations of both conventional methods and traditional AI-enabled approaches in addressing Lyapunov optimization problems.

\item We then introduce various GenAI models and provide comprehensive analysis of their potential applications within the Lyapunov optimization theoretical framework.

\item Finally, we propose a novel Lyapunov-guided framework that integrates generative diffusion models with reinforcement learning, demonstrating its effectiveness through a case study of trajectory optimization and resource allocation in UAV-based LAE networks.

\end{itemize}


   \section{Lyapunov Optimization Theory: From Conventional Methods to Traditional AI-Enabled Solutions}
     In this section, we first introduce the fundamentals of Lyapunov optimization theory and provide a brief tutorial on its applications. We then discuss the use of conventional methods and traditional AI-enabled solutions for solving Lyapunov optimization problems, highlighting their respective limitations.

    \subsection{Overview of Lyapunov Optimization Theory}
    \subsubsection{\textbf{What is Lyapunov optimization theory about}} Lyapunov optimization theory is a powerful mathematical framework designed to solve dynamic stochastic optimization problems. It decouples long-term stochastic optimization into sequential per-slot deterministic sub-problems, providing theoretical assurances for long-term system stability. Specifically, Lyapunov optimization theory uses a \emph{Lyapunov function} to consolidate all relevant queues (i.e., long-term constraints) and employs \emph{Lyapunov drift} to capture queue updates between consecutive time slots, thereby measuring system stability. The optimization objective is then defined as the \emph{penalty}, effectively balancing system stability and performance. Finally, by minimizing the \emph{Lyapunov drift-plus-penalty function} at each time slot, Lyapunov optimization theory achieves dual benefits of optimizing system performance and ensuring system stability without requiring future state information, making it particularly suitable for dynamic systems with uncertainty.

    \subsubsection{\textbf{How to use Lyapunov optimization theory}} Assume a network has $I$ queues $(Q_1(t), Q_2(t),\dots, Q_I(t))$, representing metrics such as backlogged computation and energy consumption requests at time slot $t$, which indicate how well the long-term constraints have been satisfied in previous slots. A \emph{Lyapunov function} $L(t)$ is then defined as a non-negative scalar that consolidates all relevant queues and measures their lengths. A typical quadratic \emph{Lyapunov function} is $L(t)=\frac{1}{2}\sum_{i=1}^{I}Q_i(t)^2$, which increases as the queueing system approaches instability.
    
    The \emph{Lyapunov drift}, defined as $\Delta L(t)=L(t+1)-L(t)$, is then used to capture the change in the \emph{Lyapunov function} between two consecutive time slots. Minimizing the \emph{Lyapunov drift} per time slot effectively restrains the \emph{Lyapunov function} from unbounded growth, thereby ensuring system stability. 
    
    A general \emph{Lyapunov drift-plus-penalty function} can be represented as $\Delta L(t) + Vp(t)$, where $p(t)$ is a \emph{penalty function} typically representing the objective function that reflects system performance metrics, such as task completion time and communication throughput. The parameter $V$ is a predefined non-negative weight for the penalty, indicating the importance of the objective function relative to long-term constraints at each time slot and balancing the trade-off between system performance and stability. Consequently, by minimizing the upper bound of the \emph{Lyapunov drift-plus-penalty function} at each time slot, we can achieve the dual goals of optimizing the original objective function while ensuring system stability over time (i.e., ultimately satisfying long-term constraints).

    \begin{figure*}[!t]
    \includegraphics[width=0.9\textwidth]{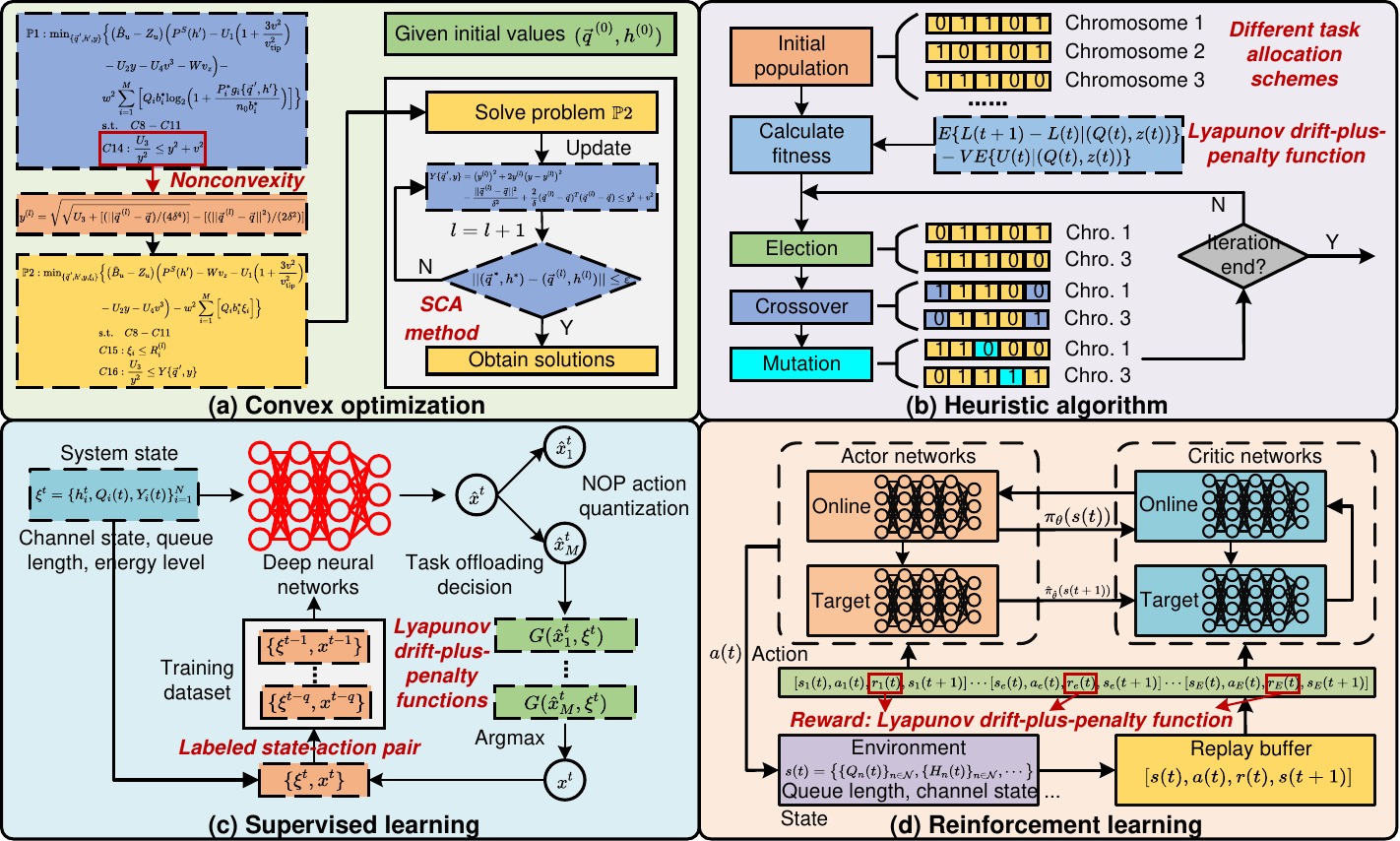}
    \centering
    \caption{An overview of conventional methods, encompassing convex optimization and heuristic algorithms, and traditional AI approaches, which include supervised learning and reinforcement learning, for addressing Lyapunov optimization problems.}
    \label{fig:Application_Examples}
    \end{figure*}
    
    \subsection{Conventional Methods for Solving Lyapunov Optimization Problems}
    \subsubsection{\textbf{From the perspective of convex optimization}} Convex optimization is a fundamental mathematical framework focused on minimizing a convex objective function over convex sets, ensuring that any local optimum is guaranteed to be globally optimal. Since Lyapunov optimization problems are typically non-convex due to complex system dynamics and queue stability constraints, researchers employ several transformation techniques to convert these non-convex problems into tractable convex formulations. For example, as shown in Fig.\ref{fig:Application_Examples}(a), the authors in~\cite{lin2024lyapunov} proposed a Lyapunov-based online control method, \emph{LYP-JRTO}, to address the trajectory planning problem for solar-powered UAV networks. Specifically, \emph{LYP-JRTO} employs the successive convex approximation (SCA) method to transform the UAV flight trajectory optimization problem into a sequence of convex, tractable problems that are solved iteratively until convergence.

    \subsubsection{\textbf{From the perspective of heuristic algorithms}} Heuristic algorithms are problem-solving methods that find near-optimal solutions by mimicking evolutionary and collective intelligence processes from nature and society. For instance, particle swarm optimization simulates social behavior to move through the solution space and simulated annealing emulates thermal process to escape local optima. As shown in Fig.\ref{fig:Application_Examples}(b), the authors in~\cite{9976258} proposed an online energy balancing strategy (OEBS) based on Lyapunov optimization for mobile crowdsensing. Specifically, OEBS employs a genetic algorithm where chromosomes (candidate solutions) form an initial population; these chromosomes evolve through two operations: crossover (for generating improved solutions) and mutation (for escaping local optima), ultimately selecting the solution that minimizes the Lyapunov drift-plus-penalty function.

    \subsection{Traditional AI Solutions for Solving Lyapunov Optimization Problems}
    \subsubsection{\textbf{From the perspective of supervised learning}} Supervised learning is a foundational machine learning approach where models learn to map input features to output labels using a labeled training dataset. For Lyapunov optimization problems, supervised learning can be employed to learn the mapping between system states and optimal control decisions through training data comprising state-action pairs. As shown in Fig.\ref{fig:Application_Examples}(c), the authors in~\cite{9449944} proposed \emph{LyDROO}, a Lyapunov-guided deep learning approach for online computation offloading in mobile edge computing networks. Specifically, \emph{LyDROO} uses Lyapunov optimization to decouple the long-term problem into per-frame subproblems, each of which is solved by using deep learning, utilizing channel gains, system queue states, and offloading decisions as labeled input-output samples to update the neural network parameters.

    
    \subsubsection{\textbf{From the perspective of reinforcement learning}} Reinforcement learning is a machine learning paradigm where agents learn optimal decisions through environmental interactions to maximize cumulative rewards. In Lyapunov optimization problems, the agent makes decisions such as resource allocation and trajectory planning based on system states (e.g., queue lengths and channel conditions), receiving rewards derived from the Lyapunov drift-plus-penalty function, and updating its policy accordingly. As shown in Fig.\ref{fig:Application_Examples}(d), the authors in~\cite{wu2020accuracy} addressed the challenge of minimizing inference service delay in deep neural network tasks while maintaining long-term accuracy requirements. To this end, they first transformed a constrained Markov decision process (MDP) into an unconstrained MDP using Lyapunov optimization techniques. Subsequently, they developed a deep deterministic policy gradient (DDPG) algorithm to optimize sampling rate adaptation and resource allocation without requiring future channel and data arrival information.
    

    \subsection{Implementation Insights} \label{subsec:lessons}
    The exploration of conventional methods and traditional AI solutions for solving Lyapunov optimization problems has yielded several valuable insights and lessons.
    
    \begin{itemize}
    \item {\textbf{Transformation Artifacts and Solution Suboptimality.}} Convex optimization methods require problem transformations that increase dimensionality and may not preserve original problem properties, potentially compromising solution quality.

    \item {\textbf{Computational Cost and Lack of Convergence Guarantees.}} Heuristic algorithms face high computational overhead, lack theoretical convergence guarantees, and show inconsistent performance due to parameter sensitivity.

    \item {\textbf{Data Requirements and Adaptation Limitations.}} Supervised learning approaches depend heavily on comprehensive training datasets and struggle to adapt to system changes post-training.

    \item {\textbf{Training Efficiency and Stability Concerns.}} Reinforcement learning methods encounter stability challenges during exploration, require lengthy training periods, and face dimensionality issues in complex systems.
    \end{itemize}

    The aforementioned limitations underscore the inherent challenges of both conventional methods and AI-enabled approaches in solving Lyapunov optimization problems, emphasizing the need to explore more effective and comprehensive solutions to advance research in Lyapunov optimization.
    
    \section{Generative AI for Solving Lyapunov Optimization Problem}

    \begin{figure*}[!t]
    \includegraphics[width=0.9\textwidth]{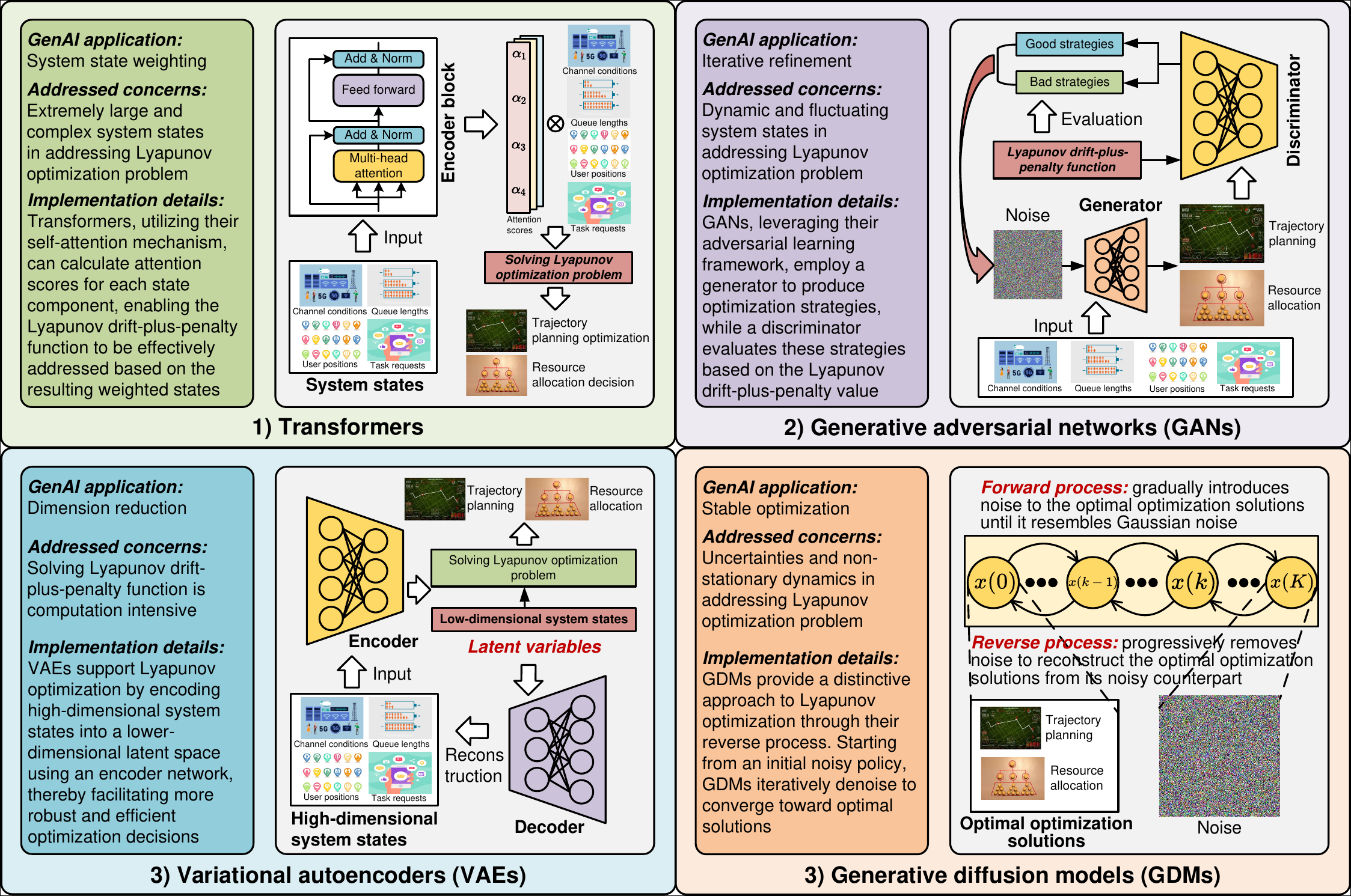}
    \centering
    \vspace{-3mm}
    \caption{A summary of the foundational architectures of key GenAI models--Transformers, generative adversarial networks, variational autoencoders, and generative diffusion models--and their potential applications in solving Lyapunov optimization problems, focusing on principles and advantages.}
    \label{fig:GenAI_Lyapunov}
    \end{figure*}
    In this section, we first overview the advantages that GenAI models can provide to Lyapunov optimization theory, followed by the comprehensive analysis of representative GenAI models and their potential contributions to advancing Lyapunov optimization.

    \subsection{Overview of Generative AI Models for Lyapunov Optimization Theory}
    GenAI has recently drawn significant attention for its unprecedented ability to automate the creation of diverse AI-generated content (AIGC), including text, audio, and images. Unlike traditional AI models, which primarily focus on discriminative tasks, pattern recognition, and decision-making, GenAI excels at generating new content by learning the underlying patterns and distributions of training data to produce novel outputs. 
    
    Beyond generating diverse digital content, GenAI models excel in solving optimization problems, thereby providing promising solutions to address the limitations and challenges presented in Section ~\ref{subsec:lessons}. For instance, variational autoencoders (VAEs) can compress high-dimensional states into low-dimensional latent variables, significantly reducing computational complexity. Generative adversarial networks (GANs) can generate substantial volumes of high-quality synthetic data with specific labels. Generative diffusion models (GDMs) can improve the balance between exploration and exploitation by adjusting the number of denoising steps.

    \subsection{Model-Specific Analysis: Working Principles and Optimization Benefits} \label{subsec:benefits}
    

    \subsubsection{\textbf{Transformers}} As shown in Fig.~\ref{fig:GenAI_Lyapunov}(a), Transformers are a deep learning architecture based on the self-attention mechanism, which has revolutionized natural language processing. By processing and weighing the importance of different parts of input sequences simultaneously, Transformers effectively capture long-range dependencies and contextual relationships. They demonstrate exceptional performance across tasks, from language translation to text generation~\cite{vaswani2017attention}. In Lyapunov optimization theory, Transformers leverage their self-attention mechanism to transform the processing of system states in Lyapunov optimization problems. Specifically, when handling complex system states--such as multiple queue lengths, channel conditions, and energy levels--the self-attention mechanism computes attention scores for each state component. Subsequently, the Lyapunov drift-plus-penalty function is addressed based on these weighted states, effectively balancing drift reduction (system stability) and penalty minimization (performance optimization).

    \subsubsection{\textbf{Generative adversarial networks (GANs)}} As shown in Fig.~\ref{fig:GenAI_Lyapunov}(b), GANs operate on a unique adversarial training principle in which two neural networks--a generator and a discriminator--compete to reach an equilibrium. The generator aims to create synthetic data that is indistinguishable from real data, while the discriminator attempts to differentiate between real and generated samples. GANs have proven particularly effective in image generation and data augmentation tasks~\cite{goodfellow2020generative}. In Lyapunov optimization theory, GANs introduce an adversarial optimization framework. Specifically, the generator network takes current system states and produces resource allocation and control decisions, while the discriminator network evaluates these decisions based on their Lyapunov drift-plus-penalty function values. Through this adversarial process, the generator learns to produce increasingly sophisticated strategies, and the discriminator aids in refining these strategies, effectively managing the trade-off between system stability and performance objectives.

    \subsubsection{\textbf{Variational autoencoders (VAEs)}} As shown in Fig.~\ref{fig:GenAI_Lyapunov}(c), VAEs utilize autoencoders to learn compact and continuous latent representations of input data. Unlike traditional autoencoders, VAEs impose a probabilistic structure on the latent space by encoding inputs into distributions rather than fixed points. Leveraging data dimensionality reduction and feature extraction, VAEs enable the generation of new samples by sampling from the learned latent distribution~\cite{kingma2019introduction}. VAEs contribute to Lyapunov optimization by mapping high-dimensional system states into a lower-dimensional latent space through an encoder network. This compressed representation captures both mean state estimates and associated uncertainties, enabling more robust optimization decisions. Meanwhile, the decoder network ensures that the latent representation retains sufficient information for accurate system state reconstruction, making the lower-dimensional optimization process more computationally efficient.
    
    \subsubsection{\textbf{Generative diffusion models (GDMs)}} As shown in Fig.~\ref{fig:GenAI_Lyapunov}(d), GDMs represent a more recent advancement in GenAI, encompassing two intertwined processes: the \emph{forward process}, which gradually introduces noise to the original data distribution until it resembles Gaussian noise, and the \emph{reverse process}, which progressively removes noise to reconstruct the original data. GDMs have gained significant attention for their stable training dynamics and high-quality generation capabilities, particularly in image synthesis tasks~\cite{ho2020denoising}. GDMs offer a unique approach to Lyapunov optimization through their reverse process. Specifically, starting from a noisy initial policy, GDMs gradually remove noise to converge toward optimal solutions. This denoising process follows a carefully designed trajectory that helps avoid local optima—a common challenge in Lyapunov optimization. This step-by-step approach not only enables more stable optimization paths but also provides an efficient way to explore the solution space.

\section{Generative Diffusion Model-Based Reinforcement Learning Framework}
In this section, we first present the rationale for adopting generative diffusion models in addressing Lyapunov optimization problems, and discuss the associated challenges. We then detail our proposed framework that integrates generative diffusion models with reinforcement learning.

\begin{figure*}[!t]
    \includegraphics[width=0.9\textwidth]{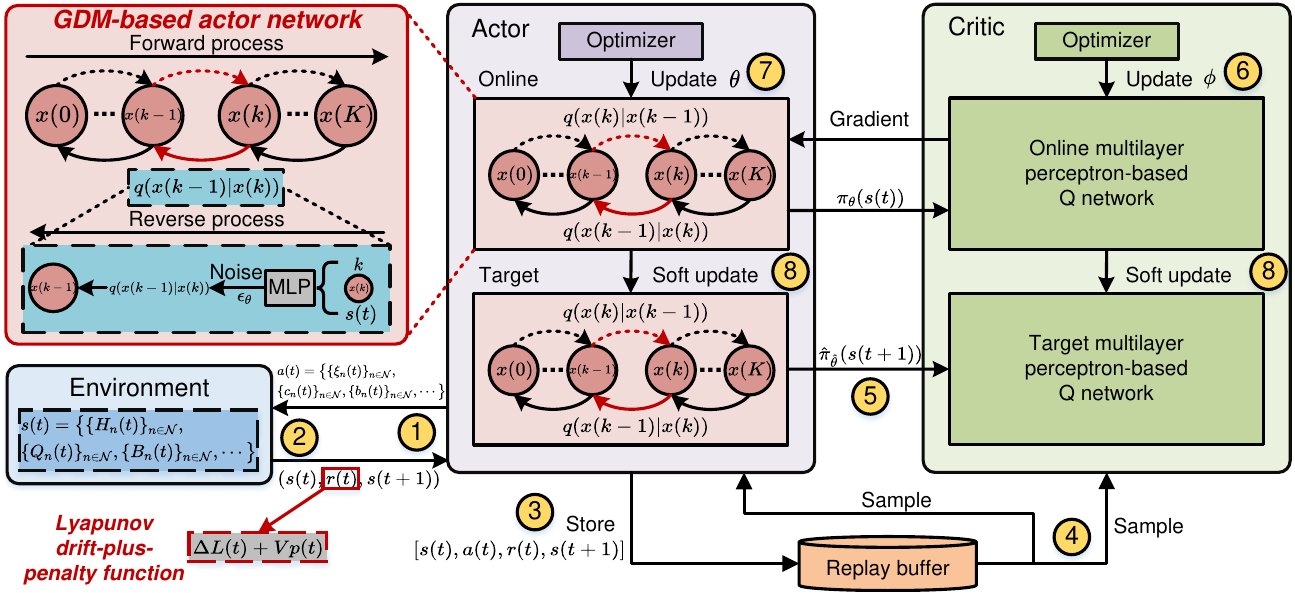}
    \centering
    \vspace{-3mm}
    \caption{The proposed GDM-based reinforcement learning framework: In \textit{step 1}, GDM generates the action $a(t)$ through the reverse process based on the current system state $s(t)$, the positional encoding of the denoising step $k$, and Gaussian noise $x(K)$. In \textit{step 2}, the environment provides feedback in the form of a reward $r(t)$ and transitions to the next state $s(t+1)$. In \textit{step 3}, the transition tuple $\langle s(t),a(t),r(t),s(t+1) \rangle$ is stored in the replay buffer for future sampling. In \textit{step 4}, transitions are randomly sampled from the replay buffer to improve the policies of both GDM-based actor networks and MLP-based critic networks. In \textit{steps 5-7}, the critic networks and actor networks are trained by minimizing the temporal difference error and maximizing the expected cumulative rewards, respectively. In \textit{step 8}, the target actor network and critic network are partially updated to stabilize the training process.}
    \label{fig:GDM_DDPG}
    \end{figure*}
    
\subsection{Motivation and Challenges}

The exploration of the potential applications of different GenAI models in Section~\ref{subsec:benefits} has demonstrated that in addition to generating diverse digital content, GenAI models can efficiently solve Lyapunov optimization problems through adaptive adjustments. Among these GenAI models, GDMs emerge as a particularly promising candidate due to their unique characteristics and advantages. 

\begin{itemize}
\item {\textbf{System Stability.}} GDMs provide a stable and controlled optimization path through their step-by-step denoising approach to solve Lyapunov optimization problems. Specifically, beginning with pure noise (random optimization solutions), GDMs iteratively remove noise to generate optimal solutions. At each denoising step, solutions are gradually refined to minimize the Lyapunov drift-plus-penalty function, avoiding sudden policy changes. This balance between stability and performance inherently aligns with the fundamental characteristics of Lyapunov optimization problems.

\item {\textbf{Local Optima Avoidance.}} GDMs progressively sample solutions to minimize the Lyapunov drift-plus-penalty function by predicting the noise that should be removed at each denoising step. During this process, additional Gaussian noise is introduced to maintain randomness and preserve the generative capabilities of the GDMs. Through this noise-guided exploration mechanism, GDMs both prevent premature convergence and avoid local optima, addressing the challenge of finding globally optimal solutions.
\end{itemize}

However, a fundamental challenge in leveraging GDMs for Lyapunov optimization problems in wireless networks lies in the inherent difficulty of obtaining optimal solutions as reference ``ground truth" data. Unlike traditional applications of GDMs such as in image generation, where original images are readily available for training, optimal solutions in wireless networks are often computationally prohibitive to obtain, especially in dynamic environments in which channel states, user demands, and network topology constantly change. This challenge is particularly critical in Lyapunov optimization, as ensuring system stability while optimizing performance objectives requires precise and reliable solution generation.
    
\subsection{The Proposed Framework}
As shown in Fig.~\ref{fig:GDM_DDPG}, we propose a novel GDM-based reinforcement learning framework\footnote{Optimization solutions in UAV-based LAE networking often involve continuous variables, such as UAV trajectory planning and resource allocation. This paper proposes a GDM-based DDPG approach, although GDMs can also be integrated with other RL frameworks like deep Q-network and soft actor-critic.} to address the challenge of requiring optimal solutions as reference data in traditional GDM applications. In this framework, optimal solutions are generated through continuous interactions with the environment, utilizing solely the reverse process. During this process, the training objective shifts from minimizing the reconstruction loss of predicted noise removal at each denoising step to maximizing the cumulative rewards in the RL framework. We next detail the framework of our proposed approach.

\begin{itemize}

\item {\textbf{State Space.}} The state space $s(t)$ typically consists of environment conditions at time slot $t$, including channel state information $H_n(t)$ that represents network conditions, virtual energy queue lengths $Q_n(t)$ of various UAVs that reflect system stability, and both computation and communication resource availability $B_n(t)$. These states provide a comprehensive view of the system's current parameters, which are fed into the GDMs to generate the corresponding optimization solutions. 

\item {\textbf{Action Space.}} The action space $a(t)$ represents optimization solutions generated through the GDM's denoising process based on observations of the current system state $s(t)$. The action space includes network control variables, i.e., UAV trajectory planning $\xi_{n}(t)$, power control levels $c_{n}(t)$, and resource allocation decisions $b_{n}(t)$, designed to minimize the Lyapunov drift-plus-penalty functions.

\item {\textbf{Reward Function.}} After executing action action $a(t)$ in response to state $s(t)$, the environment provides feedback in the form of a reward $r(t)$. This reward function is crucial, as it reflects both the Lyapunov drift and penalty terms: the drift component incentivizes queue stability by penalizing excessive queue lengths or sudden queue variations, while the penalty term incorporates performance metrics such as energy efficiency and throughput. This reward function, formulated as the negative of the Lyapunov drift-plus-penalty expression, ensures that maximizing the reward aligns with minimizing the Lyapunov objective.

\end{itemize}

Next, we present a case study to demonstrate the effectiveness of the proposed GDM-based reinforcement learning framework in addressing Lyapunov optimization problems in UAV-based LAE networking.

\section{Case Study}

In this section, we first provide a detailed description of the problem scenario and specify the parameter settings for both the network environment and learning algorithm. We then present comprehensive performance evaluations, comparing our proposed framework with both conventional and traditional AI methods.

\subsection{Scenario Description}

We consider a UAV-based LAE network scenario in the context of smart city applications, where UAVs are deployed to collect data from ground IoT devices for urban management tasks such as traffic control and congestion monitoring. In this scenario, a UAV serves as a mobile data collector, gathering information from multiple ground IoT devices. The ground devices continuously generate data requiring transmission to the UAV for real-time urban monitoring and management. Through joint optimization of UAV trajectory and bandwidth allocation ratios for IoT devices at each time slot, we aim to maximize the average uplink transmission rate from ground IoT devices over a finite time horizon. Additionally, we consider the UAV's average per-slot energy consumption constraint to maintain sustainable operation throughout the data collection process.

\begin{figure}[t!]
\includegraphics[width=.45\textwidth]{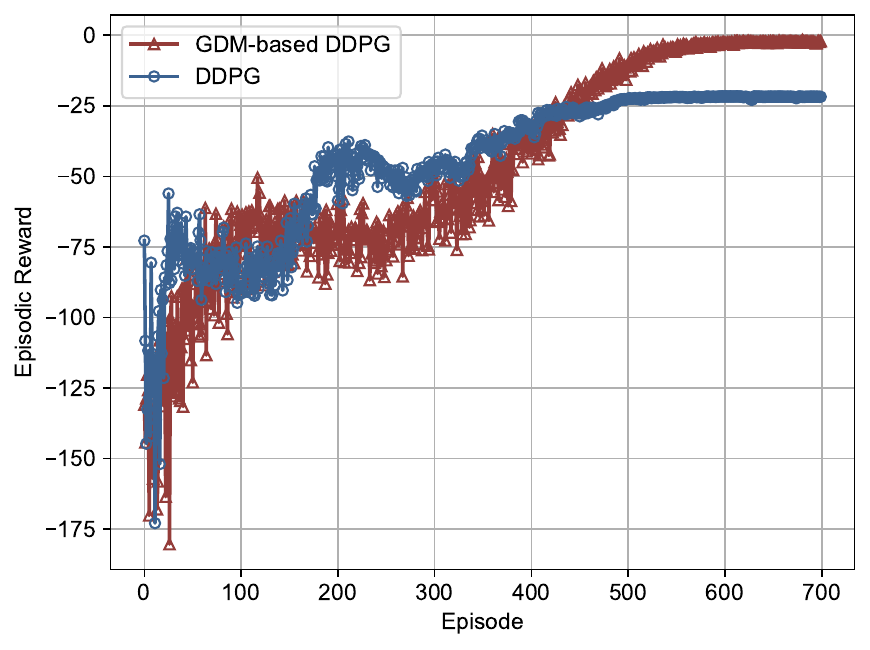}
\centering
\vspace{-1.5mm}
\caption{The training curve of the proposed GDM-based DDPG and the conventional DDPG method.}
\label{fig:D3PG_DDPG_convergence}
\vspace{-3mm}
\end{figure}

\subsection{Parameter Settings}

To evaluate the effectiveness of our proposed framework, we conduct simulation experiments for the previously considered scenario. We deploy a UAV to serve three ground IoT devices in a 600 m $\times$ 450 m rectangular area, with the UAV's initial position and destination set at [0, 0] and [600, 0], respectively. The UAV operates under an average propulsion energy constraint of 140 J and a maximum velocity of 25 m/s. The flight duration is 100 s, divided into 100 equal time slots. The communication system operates with a bandwidth of 1 MHz, and each ground IoT device transmits at 0.1 W. All experiments are conducted using PyTorch 2.0 and Python 3.8.1 on a platform equipped with an NVIDIA RTX 4090 GPU. The diffusion model employs three fully connected (FC) hidden layers with 128 neurons each to learn noise patterns. The DDPG critic networks comprise two FC hidden layers, each containing 256 neurons.

\subsection{Performance Evaluation}
    
Fig.~\ref{fig:D3PG_DDPG_convergence} illustrates the convergence behavior of the Lyapunov drift-plus-penalty function value (reward) for both our proposed GDM-based DDPG and the conventional DDPG across training episodes. The proposed GDM-based DDPG achieves an average reward of approximately 0 after 600 episodes, significantly outperforming the conventional DDPG's average reward of -25, with higher rewards indicating better system performance. Moreover, the GDM-based DDPG exhibits more stable reward convergence. The results demonstrate that GDMs in the proposed framework can iteratively reduce noise through the reverse process while maximizing rewards, thereby enhancing action sampling efficiency. The systematic denoising process not only helps avoid local optima—a common challenge in Lyapunov optimization but also provides an efficient mechanism for solution space exploration, particularly in the highly dynamic environment of UAV-based LAE networking.

\begin{figure}[t!]
\includegraphics[width=.4\textwidth]{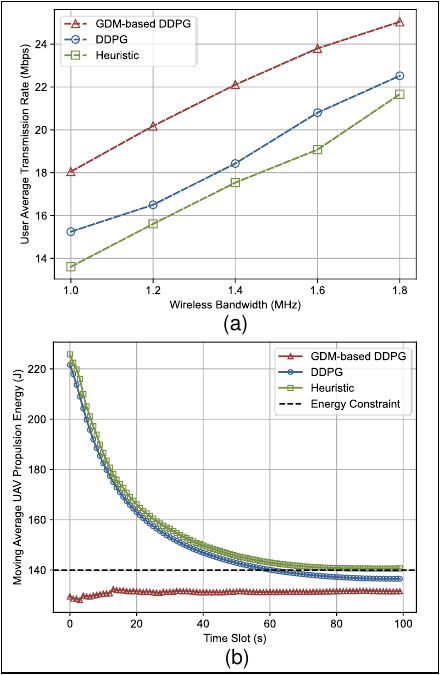}
\centering
\vspace{-1.5mm}
\caption{Performance evaluation for the proposed framework. (a) User Average Transmission Rate versus Wireless Bandwidth. (b) UAV Propulsion Energy versus Time.}
\label{fig:user_rate}
\vspace{-3mm}
\end{figure}

Fig.~\ref{fig:user_rate}(a) compares the average uplink transmission rates from ground IoT devices across different methods with increasing system bandwidth. The proposed GDM-based DDPG consistently outperforms other methods across all bandwidth levels. This superior performance demonstrates how the step-by-step denoising approach enables GDMs to adapt effectively to complex system dynamics in UAV-based LAE networking. Additionally, Fig.~\ref{fig:user_rate}(b) illustrates the UAV's moving average energy consumption across different methods during the considered flight duration. The proposed GDM-based DDPG achieves the lowest average propulsion energy consumption, demonstrating its ability to both satisfy the per-slot energy constraint for sustainable UAV operation while effectively managing the trade-off between system stability and performance objectives in the Lyapunov optimization.

\section{Future Directions}

\subsection{Generative AI for Lyapunov Optimization Theory-based Problem Transformation} 

Language models like ChatGPT can enhance problem modeling by guiding Lyapunov function design and constraint transformation strategies, offering systematic assistance in mathematical formulation.

\subsection{Generative AI for Real-time Lyapunov Optimization Adaptation}

GenAI enables dynamic adaptation of Lyapunov optimization frameworks to changing system conditions. Models like Stable Diffusion can generate adaptive virtual queue structures and drift-plus-penalty expressions, maintaining optimization effectiveness under varying conditions.

\subsection{Generative AI for Automated Algorithm Selection and Integration}

Generative models can intelligently select and integrate optimization methods based on problem characteristics. Advanced models can analyze problems to recommend appropriate methods (e.g., DDPG for trajectory optimization, convex optimization for power allocation) and generate integration frameworks.

\section{Conclusion}

In this paper, we have investigated the advantages of GenAI in addressing Lyapunov optimization problems within UAV-based LAE networking. We have introduced the fundamentals of Lyapunov optimization theory and analyzing the limitations of both conventional methods and traditional AI-enabled approaches in solving Lyapunov optimization problems. Then, we have comprehensively analyzed the potential advantages that GenAI models can provide to Lyapunov optimization theory. Subsequently, we have proposed a framework that integrates generative diffusion models with reinforcement learning to solve Lyapunov optimization problems. Through detailed case studies and simulation results, we have demonstrated the effectiveness of the proposed framework. Finally, we have highlighted several promising research directions for future work. We believe that this paper will inspire researchers to further explore GenAI models in advancing Lyapunov optimization theory.

    \bibliographystyle{IEEEtran}

    \end{document}